# Experimental Comparison of Cap and Cup Probabilistically Shaped PAM for O-Band IM/DD Transmission System

*Md Sabbir-Bin Hossain[(1,2)]\*, Georg Böcherer [(1)], Talha Rahman[(1)], Nebojša Stojanović[(1)], Patrick Schulte [(1)], Stefano Calabrò[(1)], Jinlong Wei[(1)], Christian Bluemm [(1)], Tom Wettlin[(2)], Changsong Xie[(1)], Maxim Kuschnerov [(1)], and Stephan Pachnicke[(2)]*

[(1)] Huawei Technologies Duesseldorf GmbH, Munich Research Center, 80992, Munich, Germany
[(2)] Chair of Communications, Christian-Albrechts-Universität zu Kiel, 24143 Kiel, Germany
**\***E-mail: sabbir.hossain@huawei.com

**Abstract** *For 200Gbit/s net rates, uniform PAM-4, 6 and 8 are experimentally compared against probabilistic shaped PAM-8 cap and cup variants. In back-to-back and 20km measurements, cap shaped 80GBd PAM-8 outperforms 72GBd PAM-8 and 83GBd PAM-6 by up to 3.50dB and 0.8dB in receiver sensitivity, respectively.*

**Introduction**

The demand of data traffic for data center networks (DCNs) keeps accelerating compelled by various cloud applications such as video-on-demand, gaming, cloud computing and home office, which further drives the research and development of Ethernet transceivers. Big tech companies are building mega data centers, which extend up to 10 km or beyond for fulfilling the current demand and future proofing[1]. The next generation Ethernet is expected to have transmission rates up to 800 Gbit/s or even beyond[2-4]. Due to small form factor, low foot print and cost, intensity-modulation/direct-detection (IM/DD) systems are preferred for high-speed short reach applications over coherent systems[5]. Currently, the 800G Pluggable Multi-Source Agreement is under specification with 200 Gbit/s/lane IM/DD transmission[6].

Recent development of high bandwidth (BW) electro-optical components enables transmission of data rates beyond 200 Gbit/s in C-band[7] as well as in O-band[8]. In this context, probabilistic shaping (PS) attracts interest for IM/DD systems. Considering forward-error correction (FEC) with 20% overhead (OH), 280 Gbit/s transmission has been demonstrated employing PS pulse amplitude modulation (PAM)[9]. Pure PS for the modulation format PAM-8 gives around 1 dB gain in sensitivity for a data rate beyond 200 Gbit/s[10]. Also the combination of PS and geometric shaping[11] shows around 1 dB gain in sensitivity.

In this paper, we experimentally investigate the performance of uniform PAM for modulation formats 4, 6 and 8 with symbol rates of 107, 83 and 72 GBd in the O-band, respectively. The comparison at 200 Gbit/s net bit rate considering 7% hard-decision FEC (HD-FEC) overhead (OH) also comprises for results the symbol rates of 75, 80, 85 and 90 GBd with cap shaped (Fig. 1(a)) and cup shaped (Fig. 1(b)) PAM-8. The results indicate that for 8-PAM, cap shaped always outperforms cup shaped, in contrast to the result reported in [12], where cup shaped 4-PAM was used for an IM/DD transmission. Furthermore, at 7% HD-FEC with bit error rate (BER) threshold of $3.8 \times 10^{-3}$, 80 GBd cap shaped PAM-8, with an entropy (H) of 2.6963 bit/symbol improves the receiver sensitivity by more than 3 dB compared to uniform PAM-8 and around 1dB compared to PAM-6 at the same net bit rate of 200 Gb/s. For 20 km, receiver sensitivity below -11 dBm is demonstrated.

**Cap and Cup Probabilistic Shaping**

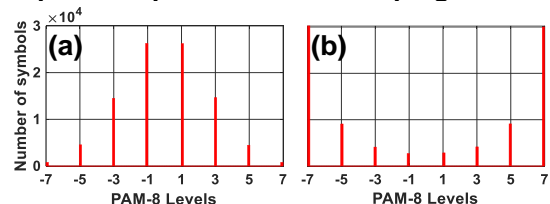

**Fig. 1**. Cap (a) and cup (b) MB shaped PAM-8.

Cap Maxwell Boltzmann (MB) distributions as in Fig. 1(a) assign probabilities $p(x) \propto e^{-vx^2}$ where $v$ is positive, i.e., the exponent is negative. Cap MB distributions achieve virtually optimal results on additive white Gaussian noise (AWGN) channels[13]. For IM/DD systems, [12] suggests to put a strong emphasis on outer points. Cup MB distributions as in Fig. 1(b) assign probabilities $p(x) \propto e^{vx^2}$ where again $v$ is positive, i.e., the exponent is positive. Both cap and cup MB distributions are symmetric around zero and therefore compatible with probabilistic amplitude shaping (PAS)[13]. The entropy of cap and cup MB distributions can be set to a desired value by choosing $v$ accordingly, for instance, larger $v$ corresponds to lower entropy.

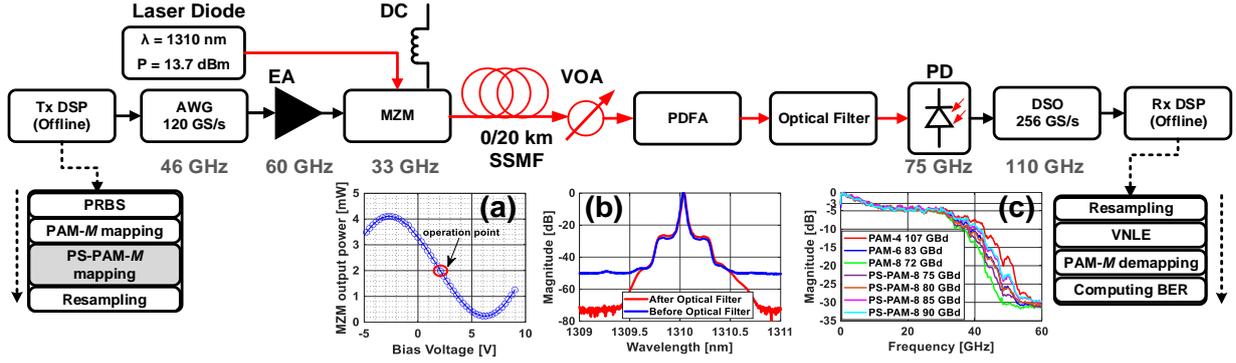

**Fig. 2:** Schematic of the IM/DD transmission system with transmitter and receiver-side offline DSP. Inset (a) shows the transfer curve of the MZM and the operation point. The optical spectrum of the signal before and after the optical filter is presented in inset (b). Inset (c) shows the power spectrum density of the received signal for different symbol rate.

**Experimental Setup and DSP**

The experimental setup is depicted in Fig. 2. Below each electro-optical component we indicate its 3-dB BW. The offline transceiver digital signal processing (DSP) is also depicted. A pseudorandom binary sequence (PRBS) was generated and mapped to *M*-levels ($M = 4$, 6 and 8) of PAM-*M*. For the modulation formats PAM-4 and 8, Gray mapping was used. For PAM-6, 5 bits were mapped on 2 symbols[14]. Similarly for PS PAM-8, random symbols were drawn from either cap or cup MB distribution with desired entropy. Considering 7% HD-FEC, targeting 200 Gbit/s net bit rate, symbol rates of 107, 83 and 72 GBd were generated for uniform PAM-4, 6 and 8, respectively. While for PS, PAM-8 with symbol rates of 75, 80, 85 and 90 GBd was generated with entropies of 2.8629, 2.6963, 2.5492 and 2.4185 bit/symbol, respectively. The generated symbol sequence was then pulse shaped with a root-raised cosine filter with a roll-off factor of 0.4 for symbol rates below 90 GBd, while roll-off factor of 0.33 and 0.12 were used for symbol rates of 90 GBd and 112 GBd, respectively.

The quantized data was then fed to an arbitrary waveform generator (AWG) with a 3-dB BW of 46 GHz, which operates at sampling rate of 120 GSa/s. To amplify the analog signal, an electrical amplifier (EA) of 60 GHz BW with 22 dB gain was used. The amplified signal was then modulated with an O-band Mach-Zehnder modulator (MZM) with 3-dB BW of 33 GHz. The MZM was biased at the quadrature point (Fig. 2(a)). Both optical back-to-back (B2B) and 20 km standard single-mode fiber (SSMF) transmissions were measured. A variable optical attenuator (VOA) was used after the fiber link, followed by a Praseodymium-doped fiber amplifier (PDFA) as pre-amplifier before a 75 GHz PIN photodetector (PD). Using the PDFA, the optical power of the PD input signal was fixed at 7 dBm. However, broadband noise generated by the PDFA was reduced by inserting an optical filter before the PD. The received electrical signal was then quantized and captured with a digital storage oscilloscope (DSO), which operates at 256 GSa/s with a 3-dB BW of 110 GHz. Data was captured for different values of received optical power (ROP) measured at the input of the PDFA adjusting by the VOA. During the capture, to avoid timing offset and jitter, the clock of the DSO was synchronized with the AWG. For offline processing at the receiver, 1 million samples were used. Synchronization was performed based on the transmitted symbols and the received signal was then resampled to 1 sample per symbol (sps). The 1 sps signal was then processed with a Volterra nonlinear equalizer (VNLE). A memory length of 311 taps was used for linear equalization to minimize the effect of reflections from the discrete component setup, copper cables and connectors. In addition to linear taps, 2$^{nd}$ and 3$^{rd}$ order kernels of VNLE with memory length of 11 were used to compensate nonlinearities introduced by EA, AWG, MZM and square-law detection. Though the VNLE was adapted in decision directed least mean square (DD-LMS) mode, the initial filter coefficients were obtained by using training sequences according to the minimum mean square error (MMSE) criterion. Afterwards symbol decision was performed on the PAM-*M* symbols followed by symbol to bits de-mapping and BER calculation.

**Transmission Results and Discussions**

The experimental results are presented in Fig. 3 in terms of pre FEC BER variation over ROP. The results are obtained with VNLE and results measured for B2B and transmission distance of 20km are shown in Fig. 3(a) and (b), respectively. As a performance measurement criterion, 7% HD-FEC. which translates to a BER threshold of $3.8 \times 10^{-3}$, is considered. Solid lines with circle markers represent uniform PAM-*M;* dash-dotted lines with pentagon markers represent cap shaped PAM-8, and dashed lines with square markers indicate cup shaped PAM-8. Furthermore, cap and cup shaped PAM-8 share the same color for the same symbol rate and the entropy.

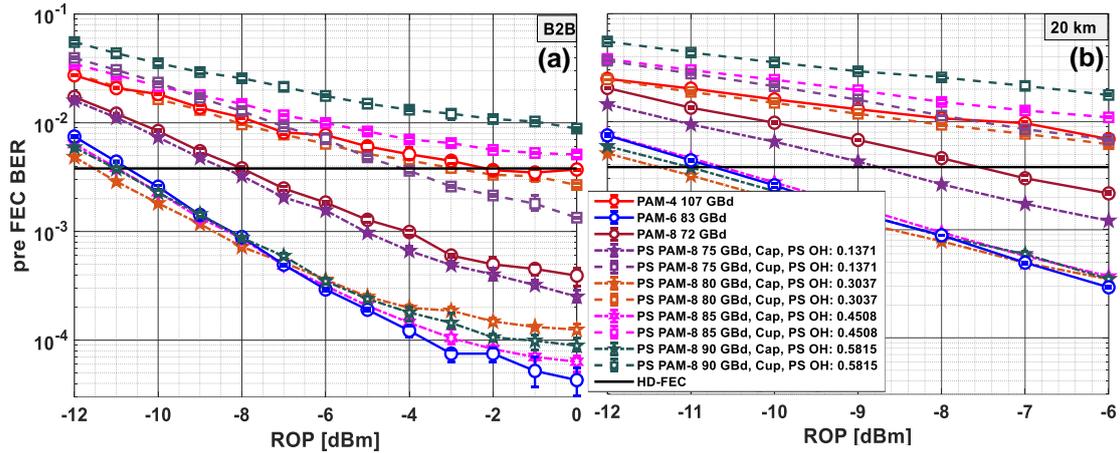

**Fig. 3:** Experimental results of uniform and probabilistic shaped PAM of different symbol rates for B2B (a) and 20 km (b).

Comparing the results for uniform PAM, PAM-6 performs the best in terms of receiver sensitivity and reached the HD-FEC threshold at -10.7 dBm. On the other hand, PAM-8 and PAM-4 reach the HD-FEC threshold at -7.94 dBm and -2.1 dBm, respectively. However, compared to uniform PAM-8, receiver sensitivity is improved by up to 0.34 dB with 75 GBd cap shaped PAM-8 with entropy 2.8629 bit/symbol. In addition, the receiver sensitivity is significantly improved by up to 3.50 dB compared to uniform PAM-8 and 0.8 dB compared to uniform PAM-6 with PAM-8 80 GBd with cap shape with an entropy of 2.6963 bit/symbol. Allowing higher BW with lower entropy does not help improving the receiver sensitivity on the contrary performance degrades due to the BW limitation of the channel.

While cap shaped PAM-8 helps to improve the receiver sensitivity to reach the HD-FEC threshold, results obtained with cup shape show an opposite trend. Cup-shaped PAM-8 with symbol rate of 75 GBd reduces the receiver sensitivity by up to 3.85 dB and by an additional dB for 80 GBd. Furthermore, symbol rates of 85 and 90 GBd with entropy of 2.5492 and 2.4185, respectively, do not reach the HD-FEC threshold. The results for 20 km transmission distance are also presented. In this case, due to attenuation (20km × 0.33 dB/km= 6.66 dB) and insertion loss, the best achievable receiver sensitivity is -6 dBm. Moreover, the performances are similar as in the B2B scenario. Uniform PAM-6, 8 and cap shaped PAM-8 reach below the HD-FEC threshold.

Eye diagrams and log histograms of PAM-8 90GBd with PS OH of 0.5815 for captured data at 0dBm are presented in Fig. 4 to explain the trend of cap and cup MB distributions. Since for cup MB PAM-8 more symbols are mapped in the outer most levels, cup MB requires a lower voltage peak-to-peak (Vpp) at the AWG output compared to cap MB PAM-8 to avoid nonlinear distortions. Consequently, the levels are kept comparatively closer and the noise distribution critically overlaps for neighbouring levels in a BW limited system like this (Fig. 2). Though the averaged eye is symmetrical, less distinguishable levels (Fig.4 (c) - blue) lead to more errors. In contrast, due to comparatively higher Vpp, the eye is skewed on the outer most levels for cap-shaped PAM-8, indicating the presence of nonlinear distortion. However, higher Vpp stretches the levels and reduces noise overlapping in between neighbouring levels, leading to better distinguishable levels (Fig. 4(c) - red), which improves the BER for cap shaped PAM-8.

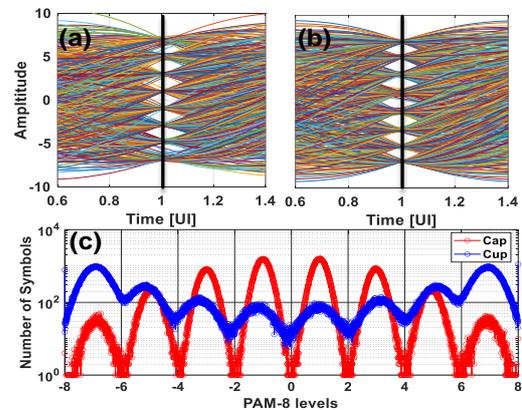

**Fig. 4:** Averaged eye diagram of (a) cap and (b) cup shaped PAM-8. Inset (c) shows the log histogram of equalized received symbols of cap and cup shaped PAM-8.

## Conclusion

We have shown an experimental comparison between uniform PAM-4, 6 and 8 with cap and cup shaped Maxwell-Boltzmann PAM-8 with different symbol rates. We targeted a net bit rate of 200 Gbit/s and allowed adequate overhead for a 7% HD-FEC. Comparing among uniform PAM formats, PAM-6 performs the best and reaches the threshold of the HD-FEC at -10.7 dBm. Cap shaped PAM-8 with symbol rate of 80 GBd outperforms PAM-6 by 0.8 dB and uniform PAM-8 by 3.50 dB both in back-to-back scenario and after 20 km transmission.